\documentclass[onecolumn]{emulateapj}
\newcommand{\HST}{\emph{HST}}
\newcommand{\Chandra}{\emph{Chandra}\ }
\slugcomment{version: 4/7/05, accepted for publication in ApJL 2005}
\begin{document}

\title{Merging Galaxies in GOODS-S: First Extragalactic Results from Keck Laser Adaptive Optics}

\author{J. Melbourne \altaffilmark{1}, S. A. Wright \altaffilmark{2}, 
M. Barczys\altaffilmark{2}, A. H. Bouchez\altaffilmark{3}, J. Chin\altaffilmark{3},
M. A. van Dam\altaffilmark{3}, S. Hartman\altaffilmark{3}, E. Johansson\altaffilmark{3}, D. C. Koo \altaffilmark{1},
R. Lafon\altaffilmark{3}, J. Larkin\altaffilmark{2}, D . Le Mignant\altaffilmark{3},
J. Lotz\altaffilmark{1}, C. E. Max\altaffilmark{1}\altaffilmark{4}, 
D. M. Pennington\altaffilmark{4}, P. J. Stomski\altaffilmark{3}, D. Summers\altaffilmark{3},
 \& P. L. Wizinowich\altaffilmark{3}
}

\altaffiltext{1} {University of California Observatories/Lick Observatory, Department of Astronomy and Astrophysics, University of California at Santa Cruz, 1156 High Street, Santa Cruz, CA 95064. jmel, lotz, koo, max@ucolick.org}
\altaffiltext{2} {Department of Physics and Astronomy, University of California, P.O. Box 951562, Los Angeles, CA 90095-1562. saw, barczysm, larkin@astro.ucla.edu}
\altaffiltext{3} {W.M. Keck Observatory, 65-1120 Mamalahoa Hwy, Kamuela, HI 96743. abouchez, jchin, mvandam, shartman, erikj, rlafon, davidl, pstomski, dsummers, peterw@keck.hawaii.edu}
\altaffiltext{4}{Institute of Geophysics and Planetary Physics, Lawrence Livermore National 
Laboratory, 7000 East Avenue, Livermore, CA 94550. max1, pennington1@llnl.gov} 
\begin{abstract}

The Center for Adaptive Optics Treasury Survey (CATS) aims to combine deep
\HST\ images in the optical with deep Keck adaptive optics (AO)  data in the
near-infrared (NIR) to study distant galaxies, AGN, and supernovae. We
recently achieved an important new milestone by securing the first 
Keck laser guide star AO image of faint galaxies.  Six
galaxies with redshifts ranging from 0.3-1.0 were targeted in one pointing
in the GOODS-S field.  Two are \Chandra Deep Field South sources, XID-56 and XID-536,
with complex morphologies suggestive of recent merger activity.  Substructures  seen in the 
NIR AO image (FWHM $\sim0.1\arcsec$), including multiple tight knots in XID-56 
and a double nucleus in XID-536, are confirmed in the optical \HST\ images.  
These structures are unresolved in the best seeing-limited ($\sim0.5\arcsec$ FWHM) 
NIR images.  Stellar population synthesis
models of the substructures indicate that XID-56 is  
a gas rich merger with a recent burst of star formation and significant amounts of dust.  
XID-536 appears to be a merger of two evolved stellar populations.
\end{abstract}


\section{Introduction}
For the past 10 years, the \emph{Hubble Space Telescope} (\HST) has provided deep, 
high-resolution imaging of galaxies in the optical, revealing 
structural parameters on kiloparsec scales to redshifts 
$z > 1$. Recently, the Keck laser guide star (LGS) adaptive optics (AO) 
system (P. Wizinowich et al. 2004) achieved similar resolution ($\sim 0.1 \arcsec$) in the NIR.
The Great Observatories Origins Deep Survey (GOODS; Giavalisco et al. 2004) 
southern field was targeted during engineering time
as part of the Center for Adaptive Optics Treasury Survey (CATS).
The CATS project aims to combine deep optical \HST\ imaging with high resolution AO data 
to study faint galaxies out to high redshift.  The high resolution imaging from optical to NIR 
enables stellar population synthesis modeling of 1-2 kpc structures to redshifts $z\sim1$ and 
beyond. The addition of Keck AO $K'$ to \HST\ $B$ to $z$ band optical photometry provides 
a wavelength range sufficient to distinguish between old stellar populations and young, 
dusty star bursts.  This work builds on previous natural guide star (NGS) AO imaging 
efforts by CATS team members, most notably Glassman et al. (2002) who 
discussed the evolution of structural parameters of galaxies in the NIR, and Steinbring et 
al. (2004), who combined \HST\ and 
NGS AO images to study the structural parameters and stellar populations of galaxies at 
redshifts $z < 1$.

This attempt is the first time that GOODS-S has been observed with the Keck AO system.  
GOODS-S is free of bright stars 
($m_R < 13$) and therefore unsuitable for NGS AO
with current high-resolution systems.  The field, however, contains 20 stars bright enough 
for tip-tilt correction with the Keck LGS AO system ($m_R < 17$). We selected a 
pointing that contains six galaxies ($K' < 22.3$)
including two \Chandra sources (Giacconi et al. 2002)
that appear in \HST\ images to be galaxy mergers. This paper will present
stellar population synthesis modeling of subcomponents within the two 
\Chandra sources, XID-56 and XID-536 (ID's from column 1 of the 1 Ms catalogue in 
Rosati et al. 2002).   The analysis indicates
that the stronger \Chandra source, XID-56, is undergoing a dusty starburst.  
The weaker \Chandra source, XID-536, appears to be a merger of two
old stellar components. 

We adopt Vega magnitudes and a flat, $h=0.7$, $\Omega_m=0.3$ cosmology throughout.

\begin{deluxetable}{ccccccc}
\tablecaption{Data for Galaxies XID-56 and XID-536 \label{table:data}}
\tablehead{\colhead{XID\tablenotemark{1}} & \colhead{R.A. (J2000)} & \colhead{Dec.} & \colhead{z} 
&  \colhead{$M_B$ \tablenotemark{2}} & \colhead{$(B-V)$\tablenotemark{2}} & \colhead{log($L_X$) [ergs/sec]}}
\startdata
56 & 03:32:13.24   &	-27:42:40.9 & 0.61 \tablenotemark{3} & -22.17 & 0.539 & 43.5 \tablenotemark{3}\\
536 & 03:32:10.76 &	-27:42:34.6   & 0.42 \tablenotemark{4} & -21.69 & 0.752 & 41.9 \tablenotemark{5}\\
\enddata
\tablenotetext{1}{Rosati et al. 2002}
\tablenotetext{2}{Photometry based on the GOODS $BVi$ and $z$ images, Giavalisco et al. 2004.  
K-correction were performed as in C. Wilmer et al. 2005, in preparation.}
\tablenotetext{3}{Szokoly et al. 2004}
\tablenotetext{4}{J. Lotz et al. 2005, in preparation}
\tablenotetext{5}{Estimated from the hard and soft x-ray fluxes in Rosati et al. 2002.}
 
\end{deluxetable}

\section{Observations and Measurements}
\subsection{AO Observations}
Observations were obtained with the Keck II (10m) telescope LGS AO system 
(Bouchez et al. 2004; P. Wizinowich et al. 2004) and the facility infrared science camera 
NIRC2, with a 1024x1024 Aladdin-3 InSb array.  On October 4, 2004 UT, the GOODS-S 
field was imaged in $K'$(2.2 microns) with the wide camera of NIRC2, yielding 
40 mas pixel scale over a field of view of 40$\arcsec$x40$\arcsec$. 
Individual frames were 120 seconds, taken using a 
5x5 dither pattern with a typical dither separation of 3$\arcsec$.  The 
field was centered 20$\arcsec$ southwest of an $m_R\sim15$ tip-tilt  star 
($\alpha=$ 03:32:13.75, $\delta=$ -27:42:13.9), and the 
laser was positioned in the middle of the field, roughly 10$\arcsec$ from each 
\Chandra source.

In addition to the science frames, $K'$ images of the tip-tilt star were obtained 
in the LGS AO mode, in order to obtain an estimate of the on-axis 
point-spread-function (PSF). These images were taken with the 10 mas pixel scale 
to fully sample the PSF.  The FWHM of the on-axis PSF is measured to be 0.06$\arcsec$.    

Image reduction was performed with typical reduction procedures of infrared imaging 
data (Glassman et al 2002).  We found that best flat field images are created by combining 
all individual AO data frames. The flat-field is created by first masking off bad pixels and 
locations of objects. Since the background at $K'$ fluctuates significantly 
due to variations in the OH night sky lines, a relative scale factor is 
calculated for each image, and this scale factor is applied to the frame 
before producing the flat-field image. The final flat is produced from the 
median of the scaled unmasked values at each pixel.
Images are reduced by dividing them by the flat and then subtracting a 
constant equal to the average remaining sky flux. Images are de-warped through 
bi-linear interpolation to remove known field distortion within the camera 
(as much as 2 pixel deviation from a true rectilinear grid).
Individual images were aligned by selecting two objects that are bright enough to be 
centroided within each frame. Once aligned, frames were combined into a final image by 
taking the median of the valid pixels at each location, yielding a total 
exposure time of 66 minutes.

\begin{figure}[h]
\centering
\includegraphics[scale=0.85,trim=0 0 0 0]{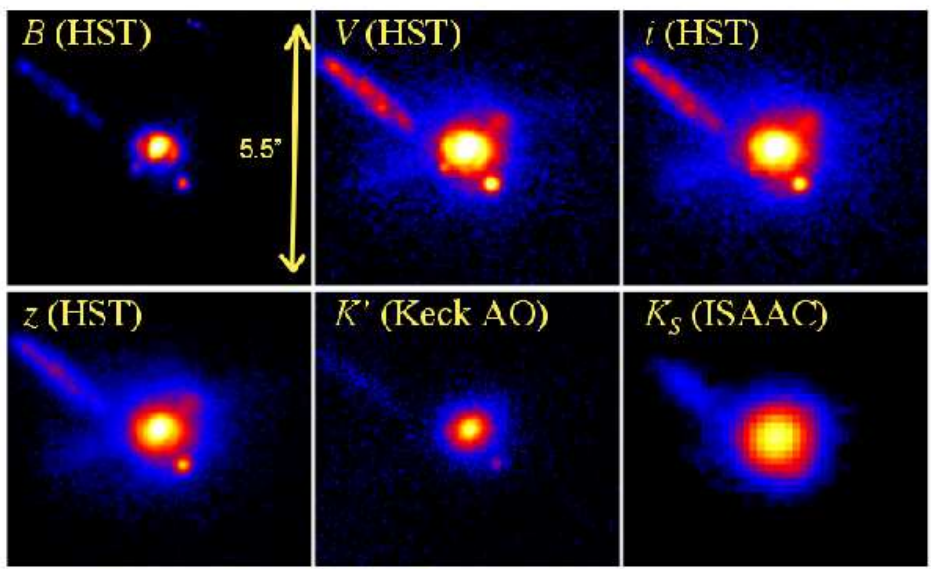}
\figcaption{\label{fig:MK} The $B, \;V,\; i,\; z$ (\HST), 
and $K'$ (Keck AO) images of galaxy XID-56.  The ISAAC $K_S$ image is also shown for comparison.  
North is up and East is to the left. The complex morphology of this galaxy is indicative of 
recent merger activity.  Several blue arc-like structures encircle
the nucleus, signs of recent star formation.  A very bright concentration to 
the south west of the galaxy appears in all five bands.  The bright
linear feature to the NE (possibly a non-thermal jet) does not point towards the nucleus of 
the bright galaxy but rather to the small object to the SW. }
\end{figure}

\subsection{Additional Data Sets}
An advantage of observing in the GOODS field is the wide variety of 
publicly available data.  This paper draws on the following: 
1) the GOODS \HST\ imaging (Giavalisco et al. 2004) in the $B,V,i$ and $z$ 
bands are used for morphology and optical photometry;
2) the 1 Ms \Chandra Deep Field South (Giacconi et al. 2002) 
observation identifies and quantifies AGN activity; and 
3) deep ISAAC imaging in the $K_S$ band (B. Vandame et al., in preparation) is used to flux 
calibrate our AO images.

Spectroscopic redshifts from Szokoly et al. (2004) for XID-56 and J. Lotz et al. 
(in preparation) for XID-536, confirm the photometric redshifts from COMBO-17 
(Wolf et al. 2004).

\subsection{Photometry and Basic Measurements }
The $K'$ AO observations are flux calibrated by matching objects in the deep ISAAC
$K_S$ images of the same region.  The color 
transformation between $K'$ and $K_S$ is on the order of 0.05 mags or smaller, a 
difference that is not large enough to affect our conclusions. 

Photometry of the \HST\ and AO images of galaxies XID-56 and XID-536 
was measured with circular apertures  
using the IDL APER program.  Sky regions were chosen at radii free of 
contamination from objects. Section 4 presents the results of small aperture 
($0.2\arcsec$ radius) photometry on the
cores of each system, while Table \ref{table:data} summarizes results of
large aperture ($4\arcsec$ radius) photometry of the galaxies.  
Within Table \ref{table:data} we list object XID (Rosati et al. 2002), 
R.A. and Dec., spectroscopic redshift (Szokoly et al. 2004; Lotz et al. 
2005, in prep), restframe $M_B$ and $(B-V)$ 
derived from the 4-band \HST\ photometry, and x-ray luminosity.  
A detailed description of the K-corrections is provided in C. Wilmer et al. (in preparation). 
The convolution between HST filter response curves and galaxy SEDs from 
Kinney et al. (1996) followed Fukugita, Shimasaku, \& Ichikawa (1995), by resampling
filters and spectra to the same dispersion (1 \AA).  Rest-frame color and magnitude  
are derived from observed color and magnitude as in Gebhardt et al. (2003).  
The x-ray luminosity of XID-56 is from Szokoly et al. (2004), while that of 
XID-536 is estimated from its measured 
hard and soft, x-ray fluxes (Rosati et al. 2002).

The photometry shows that both galaxies are luminous ($\sim$L$^\star$) with 
XID-56 0.2 magnitudes ($B-V$) bluer than XID-536.  XID-56 is a
40x stronger x-ray source than its redder counterpart 
and is classified as a Seyfert 2 galaxy based on its 
optical spectrum and x-ray emission (Szokoly et al. 2004).

\begin{figure}[h]
\centering
\includegraphics[scale=0.85]{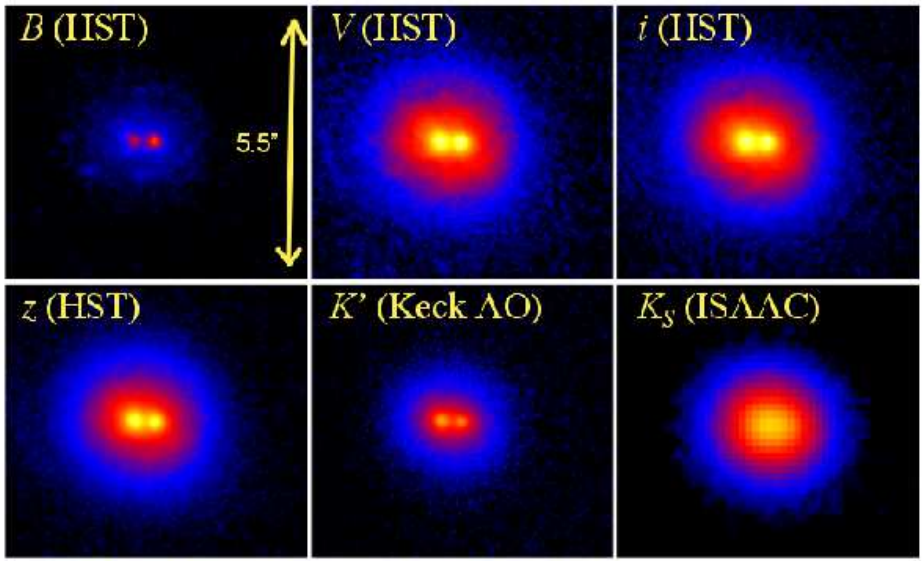}
\figcaption{\label{fig:DN} Same as Figure \ref{fig:MK} but now for  galaxy XID-536.  
The galaxy contains two distinct concentrations,
suggesting a recent merger. Both cores appear to be fairly red. The cores are unresolved in
the ISAAC image. North is up and East is to the left.}
\end{figure}
 
\section{Morphology}

\begin{figure}[h]
\centering
\includegraphics[scale=0.85, trim= 50 0 0 0]{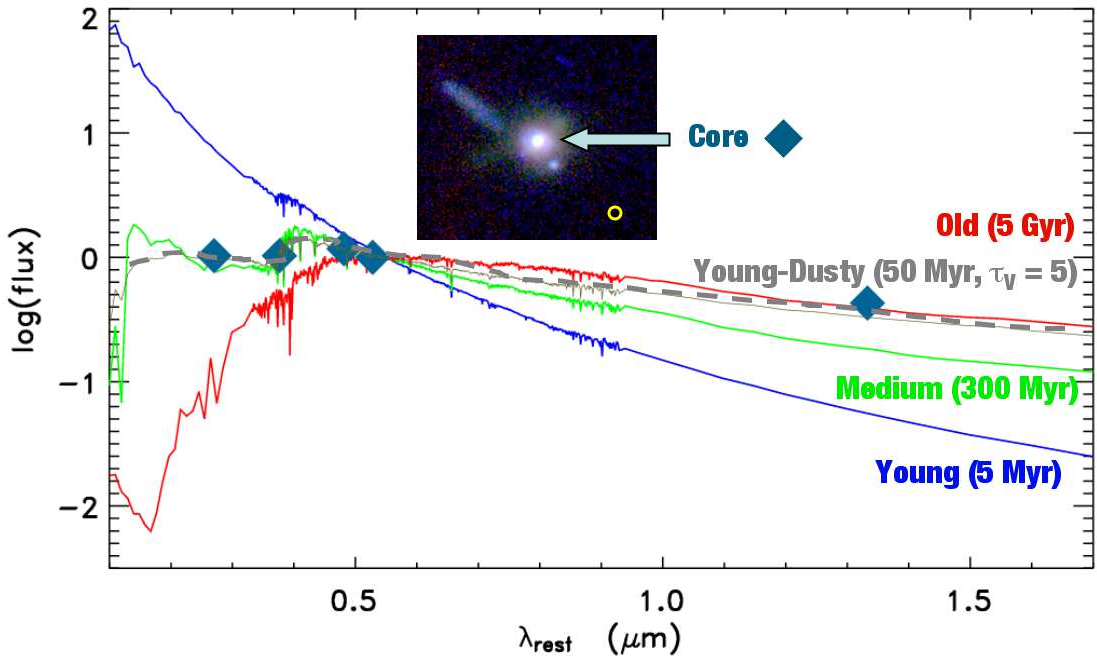}
\figcaption{\label{fig:MKsed} Bruzual and Charlot (2003) single-burst
stellar population synthesis models are plotted.  The blue spectrum is a 
young population (5 Myr), green, intermediate age (300 Myr) and red, old 
(5 Gyr).  A young burst (50 Myr) with dust (optical depth $\tau_V=5$) is also 
plotted (grey dashed).  Aperture photometry (0.2$\arcsec$ radius) of the core of 
galaxy XID-56 in the $B,V,i,z$ and $K'$ bands are shown (blue diamonds).
Photometric errors bars are the  size of the points.  The optical data (\HST) can
be fit by either the young-dusty model ($50$ Myr, $\tau_V=5$) or the intermediate
age (300 Myr) population.  However, the NIR data (Keck AO) rule out
the intermediate age and indicate that this is a young dusty population.  A three color 
image ($BiK'$) of the galaxy is shown, with North up and East to the left.  The yellow 
circle, placed in the lower right corner of the image, has a
0.2$\arcsec$ radius, equivalent to the aperture used in the photometry. 
}
\end{figure}

Both \Chandra sources have morphologies suggestive of recent merger activity.
Figure \ref{fig:MK} shows the $B,V,i,z$ (\HST) and $K'$ (Keck AO) images 
for galaxy XID-56.  Also shown is the ISAAC $K_S$ image which lacks the 
detail found in the AO image.  The morphology of this object includes: 
1)  a fairly blue core surrounded by several blue arc-like structures; 2)  a
tight knot which is bright in the optical and NIR and lies 0.83$\arcsec$ (5.5 kpc) SW of the core; and  
3)  a very blue, linear structure of length  2.5$\arcsec$ (16 kpc) that lies to the NE of the galaxy. 
VLA 1.4 GHz maps of the region (K. Kellermann 2005, private communication) 
contain a strong radio source, 
which may indicate that the linear feature is an optical counterpart to a radio jet.  While the 
radio map shows a slight extension in the direction of the linear feature the resolution 
($3.5\arcsec$)  is not high enough to confirm the nature of the feature.  Therefor we
can not rule out the possibility that the feature is a very linear tidal interaction with a smaller 
galaxy. Interestingly, the feature is not aligned with the center of the main galaxy, but rather points
towards the bright knot SW of the galaxy.  

Figure \ref{fig:DN} shows the images for galaxy XID-536.  The 
center of this galaxy is composed of two concentrations with visual sizes 
on the order of 2 kpc.   Both are fairly red, indicative of either old or dusty stellar 
populations. Because the two distinct concentrations are seen to be similar even 
in the reddest wavebands, 
they are likely to be cores of two previous galaxies in the later stages of merging, rather
than the result of a single core bifurcated by dust. The seeing limited 
ISAAC image is unable to resolve the two cores
which have a separation of 0.3$\arcsec$ or 1.7 kpc.

\section{Stellar Populations}
High resolution imaging allows us to measure the spectral energy distributions (SEDs)
of kiloparsec-scale components in each galaxy. This paper focuses on the 
cores of each system.  Figure \ref{fig:MKsed} plots the fluxes from circular
aperture photometry centered on the nucleus of XID-56 (blue diamonds), 
with an aperture radius of $0.2\arcsec$ in each band. Over-plotted are 
Bruzual and Charlot (2003; hereafter BC03) single-burst, dust-free, model spectra that span a 
range of plausible ages (5 Myr, blue; 300 Myr, green; 5Gyr, red). We use BC03 models with a 
Chabrier (2003) initial-mass function, Padova 1994 evolutionary tracks, and solar metallicity.  
Also plotted is a young (50 Myr) dusty ($\tau_V=5$) population created 
with a Charlot \& Fall (2000) dust model, with an ambient dust fraction, $\mu$, of 0.3. 
The models are normalized at a rest wavelength of 5500\AA, which corresponds to the $z$ 
band at this redshift. The optical fluxes of the core are well matched to either the young dusty 
model or the intermediate age stellar population.  The Keck AO flux in $K'$  breaks the 
degeneracy of these two models by matching the core of this galaxy to a very young and dusty model. 

To check whether light from the central AGN may be contributing to the SED, 
we generated an SED for 
an annulus with an inner radius of 0.2$\arcsec$ and an outer radius of 
0.8$\arcsec$.  The SED for this region (not plotted) is indistinguishable
within the errors from the the SED of the core.
  
\begin{figure}[h]
\centering
\includegraphics[scale=0.85,trim=40 0 0 0]{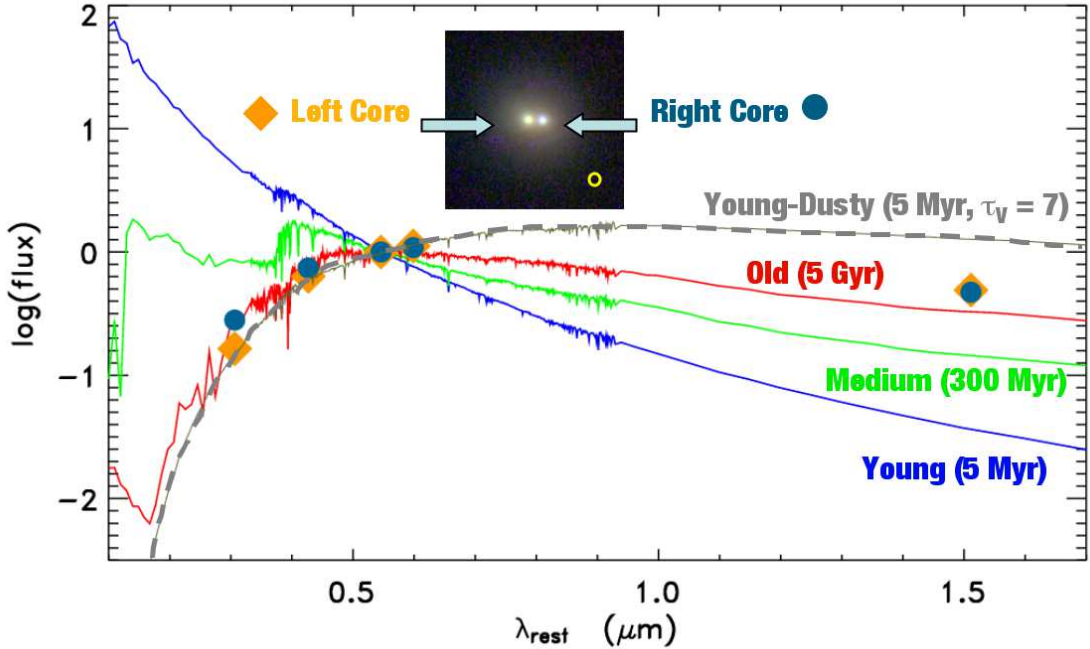}
\figcaption{\label{fig:DNsed} Same as Figure \ref{fig:MKsed}, only now we 
plot the left (E, orange diamond) and right (W, blue circle) cores of galaxy XID-536.  The 
optical data (\HST) of both cores are well fit by either the young (5 Myr), very dusty ($\tau=7$) 
population or the old stellar population.  However, the NIR data (Keck AO)
rule out the young-dusty model, indicating that the cores are both evolved 
stellar systems. The $K'$ point is slightly more luminous than
the old SED, which may indicate the presence of some dust ($\tau\sim1$) in this merger.
The right core is seen to be slightly bluer in the optical bands.}
\end{figure}

Figure \ref{fig:MKsed} shows the SEDs for the two nuclei (orange triangle and blue circle for 
E and W cores respectively) of XID-536, also using circular apertures with radii of $0.2\arcsec$.  
The two nuclei (ie. cores) have very similar SEDs, with the W core slightly  
bluer in the optical. The optical fluxes for
both cores are consistent with either the old, dust free, stellar population (5 Gyr) or
the young dusty (5 Myr, $\tau_V = 7$) model.  The Keck AO data, however, rule out 
the young dusty 
model and indicate that both cores are old, well evolved systems.  The $K'$
point lies slightly higher than the old SED model, which may indicate 
a small amount of dust in this merger ($\tau_V=1$ is consistent with the data).  It should
also be noted that the  SEDs of the two nuclei are inconsistent with the very red nuclear 
regions of luminous infrared galaxies such as Arp 220.  The SED of the core of Arp 220 continues 
to rise in the NIR (Mazzarella et al. 1992;  Shioya, Taniguchi \& Trentham 2001) 
as opposed to leveling off in the case of XID-536.
     
To ensure that the results above are insensitive to 
uncertainties in the AO PSF, we performed aperture photometry on 
the tip-tilt star images.  We compared the photometry from a $0.2\arcsec$ radius
aperture with photometry from a $0.8\arcsec$ aperture to see if we are 
missing significant light in the wings of the PSF.
The magnitude difference between the two apertures for the LGS AO PSF is $\sim0.3$ 
mags. In comparison, the magnitude difference for the \HST\ PSF's is $\sim0.2$ 
mags for all bands.  This indicates that, at most, the $K'$-band photometry should 
be shifted 0.1 mags with respect to the \HST\ photometry, an amount that does not 
alter the results of this paper.   A similar test on a point-like object in the science frame 
also indicates a $K'$ shift no greater than 0.1 mags.
      
\section{Discussion}
Hierarchical clustering models predict that the large spiral and elliptical 
galaxies of today build up from a series of mergers of smaller components
(Baugh et al. 1998; Mao, Mo \& White 1998; Somerville, Primack \& Faber 2001).
During a typical merger event, gas and dust are funneled into the center 
of the galaxy and form new stars (Mihos \& Hernquist 1996; Cox et al. 2004).  This scenario appears 
to explain the morphology of XID-56, which has a blue core and various 
blue structures indicating recent and ongoing star formation. 

One outstanding problem for hierarchical clustering models is matching 
the observed bi-modal color distribution of galaxies on the color-magnitude diagram
(Blanton et al. 2003) and specifically  creating the brightest 
red galaxies (Bell et al. 2004; Springel, Di Matteo \& Hernquist 2005). 
In current semi-analytic model prescriptions (Somerville et al. 2005, in prep) 
the star formation is either quenched before galaxies are big 
enough to become the brightest red galaxies of today, or  
residual gas inside the systems eventually becomes cool enough to form new stars, 
making them blue.  In order to make the largest red galaxies, it may be 
necessary to meet two conditions: 1) have some feedback mechanism such as 
an AGN which quenches further star formation in massive halos (Springel, Di Matteo \& 
Hernquist, 2005; Di Matteo, Springel \& Hernquist, 2005); and 2) have the 
merger be of two relatively gas free systems (Bell 
et al. 2004). While XID-536 shows signs of recent merger activity, its SED 
does not show signs of vigorous star formation.  Its luminosity ($M_B= -21.7$)
and color $(B-V = 0.75)$ are consistent with galaxies lying on the brighter  end of the red 
sequence of the color magnitude diagram.  We surmise that this is an 
example of two evolved stellar systems, where AGN activity 
has expelled or heated gas that might otherwise have been able to form new stars.  

Keck LGS AO photometry in the NIR is now possible on the same sub-arcsec scales as the 
\HST\ optical. The addition of NIR fluxes is found to be critical to distinguish among stellar population
models that otherwise match the optical SED.  In this paper, we identify two 
merging systems, one of which is unusual in that it appears to be between two 
evolved stellar populations.  Large statistically complete samples are needed to determine 
the merger rate of evolved stellar systems.  

\acknowledgments
This work has been supported in part by the NSF Science and 
Technology Center for Adaptive Optics, managed by the University of California (UC) at
Santa Cruz under the cooperative agreement No. AST-9876783.
The laser guide star adaptive optics system was funded by the W. M. Keck
Foundation. The artificial laser guide star system was developed and
integrated in a partnership between the Lawrence Livermore National Labs
(LLNL) and the W. M. Keck Observatory. The laser was integrated at Keck with
the help of Curtis Brown and Pamela Danforth. The NIRC2
near-infrared camera was developed by CalTech, UCLA and Keck
(P.I. Keith Matthews). The data presented herein
were obtained at the Keck Observatory, which is operated as a
scientific partnership among the CalTech, UC and NASA.  
This work is supported in part under the auspices of the 
US Department of Energy, National Nuclear Security Administration and by 
the LLNL under contract W-7405-Eng-48.

The authors wish to recognize and acknowledge the very significant cultural
role and reverence that the summit of Mauna Kea has always had within the
indigenous Hawaiian community.  We are most fortunate to have the
opportunity to conduct observations from this superb mountain.

\clearpage

\clearpage

\clearpage

\clearpage

\end{document}